# IS THE BRAIN-MIND QUANTUM? A THEORETICAL PROPOSAL WITH SUPPORTING EVIDENCE

Stuart Kauffman[a] and Dean Radin[b]

[a] Emeritus Professor of Biochemistry and Biophysics, University of Pennsylvania, stukauffman@gmail.com
[b] Chief Scientist, Institute of Noetic Sciences, Petaluma, CA, dradin@noetic.org

## Abstract

If all aspects of the mind-brain relationship were adequately explained by classical physics, then there would be no need to propose alternative views. But faced with possibly unresolvable puzzles like qualia and free will, other approaches are required. We propose a non-substance dualism theory, following a suggestion by Heisenberg, whereby the world consists of both ontologically real Possibles that do not obey Aristotle's law of the excluded middle, and ontologically real Actuals, that do obey the law of the excluded middle. Measurement converts Possibles into Actuals. This quantum-oriented approach solves numerous puzzles about the mind-brain relationship, but it also raises the intriguing possibility that some aspects of mind are nonlocal, and that mind plays an active role in the physical world. We suggest that the mind-brain relationship is partially quantum, and we present evidence supporting that proposition.

**Keywords**: brain-mind, quantum biology, consciousness.

## Introduction

Of the three central mysteries in science, the Origin of the Universe, the Origin of Life, and the Origin of Consciousness, the last is the most challenging. As philosopher Jerry Fodor put it in 1992, "Nobody has the slightest idea how anything material could be conscious. Nobody even knows what it would be like to have the slightest idea about how anything could be conscious."[1] Fodor's quip still holds true today.

In 1996, philosopher David Chalmers distinguished between the "easy" and "hard" problems of consciousness.[2] The "easy" problems are unraveling the physical correlates of consciousness, say electrical action potentials propagating down neural axons, or the binding of some specific molecules in the post synaptic cleft joining the axon of one neuron to the dendrites of the next. Chalmers correctly pointed out that no such knowledge would explain the mystery of being *aware* of the red color of a rose, the smell of coffee, or the pinch of pain. Philosophers call such awareness "qualia." No reasonable account of physical happenings can account for qualia, thus that is the "hard" problem.

Associated with the mystery of qualia is the problem of *will*, meaning the act of intentionally choosing and doing. The specific problem is *free will*, such that one can be held accountable for one's actions. Broadly speaking, the topic of free will is either denied or not understood.

Here we explore alternative theories about the hard problems of qualia and free will. We believe that there are testable answers, propose some of them here, and discuss supporting evidence that "mind" is at least partially quantum.

**Theories of Mind**

*Single Substance Theories*

About 400 AD, St. Augustine of Hippo, a Neoplatonist, established much of the theological teaching of the early Catholic Church. He held that human consciousness was due to a direct connection to the Mind of God.[3] In1992, the Catholic Church celebrated the 150th anniversary of Darwin's "The Origin of Species," taking the enormous step of accepting Darwin's fundamental account of and reality of biological evolution. However, the Church retained St. Augustine's conception about the source of consciousness.

About 1660, philosopher Baruch Spinoza conceived of reality as being composed of a single "substance" with two aspects, mental and physical.[4] Both were said to unfold deterministically in time, held in parallel coordination by the Mind of God. Later, a version of Spinoza's view became known as Neutral Monism, which proposes that a single substance exists which can simultaneously "be" both mental and physical. Neutral Monism is confronted with the issue of coordination between mental and physical aspects of this "neutral substance."[5]

The two principal single substance alternatives to Neutral Monism are Idealism and Physicalism. Idealism proposes that all is mind, and that matter is in some sense "congealed mind." Bishop Berkeley proposed this view.[6] Asked if a tree falling in a forest makes a sound when no one is there to hear the falling tree, Berkeley suggested that the sound existed because the falling tree was heard by God.

Physicalism proposes that all is physical, and mind is somehow an expression of physical matter. The physicalist view is the dominant philosophy today. It proposes that in sufficiently complex, classical physical neural networks, e.g. brains, consciousness spontaneously emerges. This view underlies neuroscience's focus on the neural correlates of consciousness. However, it remains unclear how neural correlates answer the hard problem of qualia and free will (as much as some have tried to provide answers).[7] For example, a classical universal computer can, in principle, be made of connected water tumblers or tin cans. There seems to be no clear reason why tin cans, in any configuration, could be conscious. Nevertheless, serious efforts concerning such classical systems have been made, most recently with an approach called "Integrated Information Theory."[8]

*Dual Substance Theories*

In 1640, René Descartes proposed two substances, *Res cogitans*, a mental substance, and *Res extensa*, extended substance.[9]  This is the famous Cartesian substance dualism, mind stuff and matter stuff.  When Descartes proposed his dualism, the Princess of Sweden asked him how mental stuff could possibly act causally on physical stuff. Descartes proposed the action occurred via the pineal gland in the brain.

This was a first step in the emergence of the mind body problem, still with us since Descartes: How does mind act causally on brain? In 1687, after Descartes, Newton's three laws of motion, the law of universal gravity, and the use of integrated differential equations yielding trajectories in phase space, founded deterministic classical physics.[10] Classical physics *is Res extensa*.

Given classical physics and its causal closure, if the mind-brain system is identical to classical physics, then the current physical state of the brain is entirely sufficient to determine the next state of the brain at the next moment. From that stance, there is nothing for mind to do. *Worse, there is no way for mind to "do it!"*  Thus, arises the familiar consequence: If mind-brain is adequately accommodated by classical physics, mind can at best be epiphenomenal and witness a world it cannot alter.  As a result, due to the determinism of classical physics, there can be no free will. Legal reasoning about intentional acts is mute, and in general attempts to deal with this problem are called "compatibilism."[11]

The proposal of an epiphenomenal mind poses overwhelming problems.  Life on Earth started about 3.7 billion years ago. Presumably, mind evolved with life; we are its progeny. We have complex experiences, qualia, and the feeling of free will, but if our minds can only passively witness the world and not alter what happens, then it is difficult to imagine why mind exists at all. More deeply, if mind is epiphenomenal, then the proposal that mind evolved from simple to complex is incoherent without an evolutionary selective advantage. We are driven to the desperate proposal that, unlike all other features of living things, our complex mind arose utterly *de-novo* at some point and was maintained at metabolic cost for no reason.

**Quantum Mechanics**

The culprit at the root of this problem is the causal closure of classical physics. We ask mind to act *causally* on the brain and body, but in classical physics all of the causes are already determined. Because of this, no form of substance dualism can work. We are left either with Idealism, Physicalism, or Neutral Monism, which is what William James and Bertrand Russell concluded early in the 20$^{th}$ Century.[12],[13]

There is another possibility. Quantum Mechanics (QM), formulated in 1927, is the most accurate theory in physics, confirmed to at least 13 decimal places. QM has transformed our understanding of reality by replacing Newtonian determinism with

Page 3 of 24

indeterminism. It breaks the causal closure of classical physics, creating new possibilities for Mind to enter.

Until about two decades ago, the proposal that quantum effects could occur in the warm wet environment of living cells was dismissed due to assumptions about environmental decoherence, which occurs on a femtosecond time scale. Today, increasing evidence for quantum effects have been found in diverse areas of biology, from bird navigation,[14], to olfaction,[15] to photosynthesis.[16]

We will base our discussion of the evidence for quantum mind in part on a novel interpretation of quantum mechanics.

### Res Potentia and Res Extensa linked by measurement

In 1958, Heisenberg proposed that the quantum state concerns "potentia, ghost like, between an idea and a reality."[17] We shall build upon this concept.[18],[19] Consider that the concept of a superposition, e.g. "the cat is simultaneously alive and dead," is a logical contradiction, so superpositions do not obey Aristotle's law of the excluded middle. On the other hand, the results of quantum measurement, e.g. "the spot is here on the screen," does obey the law of the excluded middle. C. S. Pierce pointed out that *actuals* obey the law of the excluded middle, but *possibles* do not. Consider that "the cat is simultaneously possibly alive and possibly dead" is not a contradiction.

Following Heisenberg then, we propose a *non-substance dualism*. In this view, the world consists of both ontologically real Possibles, *Res potentia*, that do not obey the law of the excluded middle, and ontologically real Actuals, *Res extensa*, that do obey the law of the excluded middle. Measurement converts Possibles into Actuals. Such a "becoming" is not deductive. The "X is possible" of *Res potentia* does not entail the "X is actual "of *Res extensa*. Indeed, no deductive mechanism has been found since the foundations of QM, suggesting that no such mechanism exists.

The "*Res potentia* and *Res extensa* linked by measurement" interpretation explains five mysteries of quantum mechanics: spatial non-locality, which-way information, null measurements, why measurement of one of N entangled variables instantaneously alters the amplitudes of the remaining N -1 variables, and why there are "no facts of the matter" between measurements, [18,19, 20]. No other current interpretation of QM explains these five mysteries with a single hypothesis. Thus, it is reasonable to take seriously the hypothesis that measurement converts ontologically real Possibles into ontologically real Actuals.[18,19]

*Res potentia* and *Res extensia* is not a substance dualism, so it does not inherit the Mind Body problem. We are no longer limited to a choice of Descartes, Idealism, Materialism, or Neutral Monism. With respect to Neutral Monism, since Spinoza it has had the problem of accounting for the coordination in time between mind and body. For Spinoza, this was accomplished by God.

We provide a different hypothesis: Mind *acausally* mediates Actualization of Potentia.[20,21] That is, we propose that a partially quantum mind-body system allows a mind to have acausal consequences for the classical brain. In this case, mind is not merely epiphenomenal, *and therefore mind can have evolved.* Moreover, if mind mediates the actualization of potentia, this automatically accounts for the temporal correlation of mind and body.

The hypothesis that "mind" acausally mediates Actualization of Potentia is a scientific hypothesis, subject to test. We discuss experimental evidence supporting this prediction below.[20,22–25]

### Experimental Predictions of Quantum Mind

Spatial nonlocality is fully established.[26] There is also evidence for temporal nonlocality,[23] e.g., photons that never existed at the same time can be entangled.[24] If mind-body is partially quantum, then certain kinds of nonlocal experiences should be viewed as physically plausible. If such experiences are ultimately based on entanglement, and presumably entanglement in living systems is fleeting, then these experiences would likely be fragile and require large-scale experiments, or meta-analyses of many independent replications, to detect the effects under well-controlled conditions.

One could predict that two types of nonlocal experiences would be reported: The mind would have the capacity to extend beyond the mind-brain system, and the act of observing a distant physical system would, to some degree, directly influence the behavior of that system. Such effects ought to occasionally result in experiences where minds interact with other minds, where minds perceive hidden or distant objects or events, and where minds directly influence aspects of the physical world. Such experiences have been recorded in every culture throughout history and at every educational level.[29] They are even reported by a high proportion (over 90%) of randomly sampled contemporary scientists and engineers.[30]

Such experiences, dubbed psychic or "psi" (first letter in Greek of the word *psyche*) for short, are so common that over the years many terms have been proposed to describe them. The most common in today's vernacular are *telepathy* for mind to mind interactions, *clairvoyance* for perceptions through space, *precognition* for perceptions through time, and *psychokinesis* for mental influence of physical objects. Use of these terms does not imply that the underlying phenomena are different; they are just labels used to describe the way the experiences seem to manifest.

The existence of psi effects has faced robust ridicule by some scientists because they imagine that such phenomena are theoretically impossible.[31] Ridicule would be justified if the world only consisted of classical physics, but of course, that is not the case. If these phenomena were well confirmed in properly designed and executed experiments, and if they were reframed as theoretically plausible, then scientists should take the evidence for these effects more seriously because the implications go to the heart of quantum mechanics, the physical correlates of consciousness, and beyond.

# Evidence

The volume of anecdotal reports of psi experiences suggests that something interesting is going on, but scientific credibility rests upon independent replications of controlled laboratory experiments published in peer-reviewed journals. Such evidence is available because psi effects have been systematically studied using the tools and techniques of science starting in the late 1800s. The discipline adopted the name "parapsychology" in the early decades of the 20th century, and in 1969 the American Association for the Advancement of Science (AAAS) elected the professional organization of scientists and scholars interested in psi, called the Parapsychological Association, as one of its affiliated societies.

To investigate each of the four classes of psi phenomena, multiple experimental protocols have been developed. Some have become standard laboratory designs replicated dozens to hundreds of times in labs around the world. These classes of experiments have been meta-analyzed to assess whether the reported effects are repeatable, to provide estimates of effect sizes, to study how variations in experimental quality might influence effect sizes, and to assess possible biases due to selective reporting.

A 2018 review of this evidence, appearing in *American Psychologist*, the flagship journal of the American Psychological Association, surveyed the results of over a thousand published psi experiments in 11 categories. The conclusion of that review was unambiguous:

> The evidence provides cumulative support for the reality of psi, which cannot be readily explained away by the quality of the studies, fraud, selective reporting, experimental or analytical incompetence, or other frequent criticisms. The evidence for psi is comparable to that for established phenomena in psychology and other disciplines.[32], p1

In 2016, a similar conclusion was reached by the President of the American Statistical Association. She wrote:

> For many years I have worked with researchers doing very careful work in [parapsychology], including a year that I spent full-time working on a classified project for the United States government, to see if we could use these abilities for intelligence gathering during the Cold War. . . At the end of that project I wrote a report for Congress, stating what I still think is true. The data in support of precognition and possibly other related phenomena are quite strong statistically and would be widely accepted if it pertained to something more mundane. Yet, most scientists reject the possible reality of these abilities without ever looking at data! [33], p. 1379

That last comment may seem like an exaggeration, but the reality is that leading skeptics today unapologetically insist that "there is no good reason to consider the

data."[34],p.8 That statement appeared in an article published in response to the 2018 article that appeared in *American Psychologist*. The authors argued:

> If the physicalist-materialist framework of modern science is correct within the bounds of demonstrability and theoretical coherency—and everything that has been learned through science says that it is—the fact that claimed parapsychological phenomena are so grossly inconsistent with that framework suggests that they are all but impossible and that the claims made by proponents cannot be true.[34], p2

This is a common argument, but it assumes that classical physics is completely sufficient to understand the mind-brain relationship, and that quantum effects cannot have any influence on living systems. Based on those mistaken beliefs, critics feel justified in ignoring the relevant empirical database. The history of science has repeatedly shown that this sort of hubris is a poor bet on what the accepted worldview of tomorrow will bring, especially given that the world is no longer the physicalist–materialist framework of classical physics. The world is quantum, and by virtue of spatial and temporal non-locality, psi phenomena are not only physically possible, but are expected to occur if mind is, in part, quantum. The idea of "quantum consciousness" in warm, wet, noisy brains is now a legitimate topic of discussion,[35,36] and the consequences of QM effects on nonlocal properties of consciousness have been proposed, largely by physicists, since the 1970s.[37–42]

Here we briefly review three categories of relevant experiments, one on telepathy, a second on precognition, and a third on psychokinesis. Each category is described in hundreds of articles in the peer reviewed literature. A comprehensive review of all of the relevant empirical evidence is beyond the scope of this paper.

**Telepathy**

*Conscious impressions*

Telepathy postulates the existence of nonlocal mental connections between minds isolated by distance or shielding. Over the past four decades, the technique most frequently used to study these connections is called the ganzfeld experiment, where *ganzfeld* is a German word meaning "whole field." The method involves exposing a "receiver" of telepathic information to a mild form of unpatterned sensory stimulation. In this experiment, the receiver relaxes in a comfortable, reclining chair, the experimenter tapes halved Ping-Pong balls over the receiver's eyes, and then asks her to wear headphones that are playing pink noise, a whooshing sound like a deep-throated waterfall.[43] Then the experimenter shines a red light on her face, and she is asked to keep her eyes open. Resting in this condition for a few minutes encourages the mind to slip into a dream-like hypnogogic state, which is thought to enhance the ability to pay attention to subtle mental impressions.

After the receiver relaxes in this reverie state for about 15 minutes, she is asked to speak aloud anything that comes to mind over the next 20 minutes while the "sender," who is strictly isolated from the receiver by shielding and distance, mentally sends her a target image. The image is one randomly selected photograph (or video clip) out of a randomly selected pool of 4 photographs (or videos), each of which is as different from one another as possible. The typical ganzfeld experiment will have 20 to 40 such target pools prepared in advance. None of the photos used in the experiment are known to the sender or receiver, and no one who directly interacts with either pair is allowed to know the identity of the randomly selected target pool or photo in a given test session.

After the 20-minute sending period ends, the receiver is taken out of the ganzfeld condition, shown four photographs, and asked which one best matches her impressions of what the sender was mentally "transmitting." By chance, assuming telepathy did not exist, she would select the correct target on average 1 in 4 times, for a 25% "hit rate." If after repeated trials with many sender-receiver pairs the overall hit rate were sufficiently above chance expectation, and experimental controls were firmly in place, then this would provide evidence that information was somehow communicated from sender to receiver. We may note here that because the existence of telepathy, and psi in general, are regarded as controversial, to prevent any form of conventional information from passing between sender and receiver, these studies follow strict security procedures that were explicitly established in consultation with both skeptics and stage magicians (because of their familiarity with deceptive practices) to close all known loopholes.[43–45]

From 1974 to 2018, the combined ganzfeld database contained 117 studies. Of those, studies using targets sets with 4 possible targets included 3,885 test sessions, resulting in 1,188 hits, corresponding to a 30.6% hit rate.[46,47] With chance at 25%, this excess hit rate is 8.1 sigma above chance expectation ($p = 5.6 \times 10^{-16}$). Analysis of these studies showed that similar effect sizes were reported by independent labs, that the results were not affected by variations in experimental quality, and that selective reporting biases could not explain away the results. The Bayes Factors (BF) associated with the last 108 more recently published ganzfeld telepathy studies was 18.8 million in favor of H1 (i.e., evidence favoring telepathy). Given that BF > 100 is considered "decisive" evidence, this outcome far exceeds the "exceptional evidence" said to be required of exceptional claims.[48,49] By comparison, in particle physics experiments effects resulting in 5 or more sigma are considered experimental "discoveries."[50] Of course, an important difference between particle physics and psi studies is that with the latter the evidence is typically evaluated via meta-analysis, which integrates results over many conceptually similar replications, whereas in particle physics an experiment might rely upon a single study that generated vast amounts of data.

The modest 5% advantage over chance expectation in the ganzfeld telepathy studies suggests that rudimentary forms of telepathy are widely distributed among the general population. We know this because the majority of participants in these studies were unselected, often just college students participating in an experiment to gain credit for their psychology courses. By contrast, in a subset of these studies where participants were selected based on their prior reports of telepathic experiences, maintaining an active meditative practice,[51] engaged in creative pursuits,[52] and/or having strong belief in psi,[53] the hit rate was a more robust 40.1%, some 6.2 sigma above chance expectation (p = 2.8 x $10^{-10}$).[54]

If some people have a natural predisposition for these abilities, possibly due to genetic differences, then such capacities can be subject to natural selection and may indeed have evolved. This in turn opens the door to heritability analyses in any capacity for telepathy or other psi phenomena. Confirmation of heritability would be powerful evidence for both the reality of psi phenomena, and for a role that these aspects of mind may have had in its evolution.

Preliminary evidence for psi heritability already exists in a survey among families in Northern Scotland who claimed a precognitive ability dubbed "second sight,"[55] and also in a recent genomics study, where differences were identified in a comparison of DNA obtained from vetted psychics from psychic families versus matched non-psychic controls.[56]

Critics may argue that telepathy experiments were conducted only by "believers," which raises doubts about the credibility of the published results. What would happen if hard-core "disbelievers" conducted these same studies? In 2005, psychologists who explicitly disavowed belief in any form of psychic phenomenon attempted to replicate the ganzfeld telepathy experiment using the methods described above. In their published report, they wrote, to their chagrin:

> After eight studies, we had an overall hit rate of 32% (which agrees with the positive meta-analyses) and, in fact, our hit rate was also statistically significant. [57],p298.

*Electrocortical evidence*

Another way that telepathy has been tested takes advantage of the neural correlates of consciousness. These experiments explored if electrocortical activity in the brains of isolated pairs of people would show significant correlations when one member of the pair was exposed to audio, light flashes, or other stimuli known to generate event-related potentials. One of the earliest experiments of this type, involving sets of identical twins, was published with positive results in 1965 in *Science*.[58] Another was reported, again with positive results, in 1974 in *Nature*.[59] Over a dozen other experiments have been reported using similar protocols, including studies with positive outcomes conducted using functional MRI.[60,61],[62] Formal meta-analyses have not been attempted on this class of experiments because the broad range of methods used makes it difficult to uniformly compare them in quantitative terms. But a case can be made that the preponderance of these neurophysiological studies are in agreement with the results of the ganzfeld telepathy studies.[63]

*Autonomic nervous system evidence*

Another approach to investigating telepathy has been based on the observation that both the ganzfeld and the brain-to-brain correlations studies seemed to involve the central nervous system, so perhaps similar effects would appear in the autonomic nervous system (ANS). Experiments examining the predicted correlations between isolated monozygotic twins, in parameters such as electrodermal activity and variations in heart rate, reported positive correlations in two separate experiments.[64,65] While those studies are intriguing, a larger class of ANS experiments has been tested within a paradigm based on the commonly reported "feeling of being stared at." The typical protocol in such studies isolates two people, where a "receiver" is monitored for changes in electrodermal or other ANS reactions as they relax quietly, while a distant "sender" periodically views their image over a live closed-circuit video channel. The sender is asked to mentally "contact" or simply gaze at the receiver when the image appears, and to withdraw their attention when the image disappears. The hypothesis is that under conditions where ordinary sensory clues are strictly excluded, the receiver will unconsciously sense when they are being observed by the distant sender, and their nervous system will react accordingly.

A meta-analysis of these experiments retrieved 36 high-quality studies consisting of 1,015 individual sessions. The overall result was 3 sigma ($p = 0.001$) above chance expectation.[66] A second meta-analysis of 15 additional studies using similar protocols, and involving 379 sessions, resulted in an overall 2 sigma ($p = 0.01$) outcome. Examination of the distribution of effect sizes in these studies revealed no selective reporting biases that could have favored a positive outcome, and the correlation between study quality and effect size was not significant. The author of the meta-analysis cautiously concluded that "the existence of some anomaly related to distant intentions cannot be ruled out".[66],p. 245

### Precognition

Precognition postulates that perception transcends the usual unidirectional constraints of time. Three primary classes of precognition studies have been reported. One involves selection of randomly determined future targets, and two involve unconscious responses to future stimuli.

*Forced-choice tasks*

In the first case, a meta-analysis of forced-choice precognition experiments conducted between 1935 and 1987 (the date of the published meta-analysis) found 309 studies reported in 113 publications reported by 62 different principal investigators.[67] Over 50,000 participants contributed nearly two million trials in these studies. The experimental design asked participants to guess which of a limited set of possible targets would be randomly selected in the future. The results of each experiment were evaluated as the proportion of observed hits over many repeated trials, as compared to chance expectation. The result was a small positive effect size, but given the large sample size the effect was 11 sigma above chance ($p = 6.3 \times 10^{-25}$). The effect size remained about the same over a half-century of replications while experimental quality systematically improved, and selective reporting biases were deemed incapable of reducing the overall results to null.

*Physiological tasks*

While the forced-choice tests provided intriguing evidence in favor of precognition, the small effect size required very large datasets to yield significant results. This led researchers to explore more sensitive measures, including unconscious physiological responses. One of the more successful approaches did not require the participant to consciously guess the future (pre-*cognition*), but rather to unconsciously "feel" the future (pre-*sentiment*). This protocol was based on a laboratory simulation of what is reported in real life as spontaneous gut feelings or hunches about unpredictable future events.[68]

In a presentiment experiment, the participant's skin conductance, heart rate, pupil dilation, EEG, or some other physiological measure is recorded before, during, and after exposure to a series of randomly selected stimuli.[69] The stimuli might be photographs with different degrees of emotionality and valence, or the presence or absence of light flashes, audio tones, or electrical shocks. The hypothesis is that a few seconds prior to exposure, the physiological measure would respond differentially to the upcoming calm versus emotional or startling future events.

The first published meta-analysis of these studies retrieved 26 relevant publications from 1978 to 2010.[70] The results showed a significant effect ranging from 5.3 sigma ($p < 5 \times 10^{-8}$) to 6.9 sigma ($p < 2 \times 10^{-12}$), depending on whether a random effects or fixed effects model was employed. Higher quality experiments resulted in larger effect sizes and greater levels of significance than lower quality studies, and selective reporting was deemed insufficient to explain the results.

A more recent update found 27 new or previously unretrieved presentiment experiments conducted from 2008 to 2018. Results were compared between peer and non-peer-reviewed publications, and in preregistered versus non-preregistered studies. The results showed that peer-reviewed studies were associated with an overall 7.6 sigma outcome ($p < 10^{-14}$), whereas the non-peer-reviewed studies were associated with $p < 0.02$. Fourteen preregistered tests resulted in a 3.3 sigma effect ($p < 4 \times 10^{-4}$), whereas 22 non-preregistered tests resulted in $p < 0.007$. There was no evidence of publication bias, the results of both frequentist and Bayesian analyses converged to similar significant results, and there were no significant differences between effects found in peer and non-peer reviewed experiments. The authors' conclusion: "This phenomenon may hence be considered among the more reliable within those covered under the umbrella term 'psi'."[71]<sup>;p. 7</sup>

*Behavioral tasks*

A third category of evidence for precognition involves reversing a standard behavioral task in time. Consider an experiment in which a participant is asked to select which of two photographs is more likeable. The pair of photographs in question is previously judged by many people to be equally likely, so the question studied in this experiment is what causes someone to select one of those images over the other. Many previous experiments have demonstrated that people prefer images that they have seen before. This "mere exposure" effect is one of the reasons why the same products are constantly advertised on television. The more you are exposed to ads for, say, Coca Cola, the more it is regarded as the favored choice.[72]

The mere exposure effect is so strong that even if one of the two images were presented subliminally before the selection were made, the unconscious priming is still sufficient to cause that image to be preferred.[73] To turn this phenomenon into a precognition experiment, rather than subliminally present one of the images *before* the selection is made, it is presented *afterwards*. The hypothesis is that future exposure to one of two equally likely images will result in a mere exposure preference for that image in the present.

This experimental paradigm was first published in 2011.[74] It reported nine experiments with overall highly successful results. A replication package was made available to any qualified researcher who wished to replicate the study. By 2016, some 90 replications were available and a meta-analysis was published.[75] Studies were reported from 33 laboratories in 14 countries, and the overall results were 6.4 sigma above chance ($p = 1.2 \times 10^{-10}$). The Bayes Factor was $5.1 \times 10^9$, far beyond the threshold for decisive evidence. When the originally reported nine experiments were excluded from the meta-analysis, the overall result was reduced to 4.2 sigma ($p = 1.1 \times 10^{-5}$) and the BF value to 3,853, still more than decisive evidence.

**Psychokinesis**

*Random physical systems as targets of mental intention*

In 1627, Sir Francis Bacon published *Sylva Sylvarum*, a tome that argued for the value of empiricism.[76] In that work, Bacon proposed that the "force of imagination," by which he meant intention or will, could be studied by focusing the mind on objects he described as having "the lightest and easiest motions," including the "casting of dice." That suggestion presaged by 300 years the use of tossed dice in experiments designed to test the effects of intention on the behavior of random physical systems.

Systematic scientific tests of psychokinetic effects on dice began in 1935.[77] A meta-analysis published in 1989 found 73 relevant publications, representing the efforts of 52 investigators from 1935 to 1987.[78] Over that half-century, some 2,500 people attempted to mentally influence 2.6 million dice-throws in 148 experiments, and 150,000 dice-throws in 31 control studies where mental influence was not applied. The number of dice tossed in a single throw ranged from 1 to 96.

The overall effect size was small, but the overall results were 19 sigma over chance expectation ($p \approx 10^{-80}$). A subset of 59 studies that used dice designed to ensure that each die face had the same probability of landing face-up (so-called "perfect dice," like those used in casino games), and where the experiments reported homogeneous effect sizes, resulted in a more modest 2.6 sigma result (p = 0.005). The results of control experiments were uniformly within chance expectation. Variations in design quality and the possibility of selective reporting could not explain away these results.

Because dice experiments tended to be labor-intensive, as electronic circuitry advanced more efficient approaches were explored. Perhaps the best known of these efforts were published by physicist Helmut Schmidt, who developed automated electronic random number generators (RNG), where the output consisted of sequences of truly random bits (0s and 1s).[79]

In the earliest RNGs, randomness was based on latencies between emissions of radioactive particles and data were recorded on paper tape. As technology evolved, later versions were based on noise from electron tunneling in Zener diodes and data were recorded digitally. From the beginning, RNGs were designed to be impervious to all known environmental influences encountered in normal laboratory conditions (e.g., variations in temperature, electromagnetic fields, etc.), and the outputs were further "whitened" by passing the random bits streams through logical filters (e.g., exclusive-OR gates). Devices used in these tests passed industry standard randomness testing suites and became a standard tool for conducting RNG experiments.

Schmidt reported highly significant effects in many of his experiments,[80] encouraging some researchers to try to replicate his results, and others to propose theoretical models based on the role of observation in quantum mechanics.[79,81] One of the more systematic and rigorous replications was performed over a period of 12 years at Princeton University's Engineering Anomalies Research laboratory.[37] That study confirmed the effects reported by Schmidt.[82]

The first meta-analysis of these studies, published in 1989, retrieved 152 publications, describing 597 experimental and 235 control studies reported by 68 different principal investigators.[83] Any experiment using an RNG as the target of mind-matter interaction was included in that analysis. The results showed a 6.8 sigma effect (p = 5.23 x $10^{-12}$) in the experimental data, null results in the controls, no significant effects due to selective reporting, and null correlations with assessed experimental quality.

A second meta-analysis published in 2006 examined a subset of the relevant literature that involved only studies using very similar designs and protocols.[84] It included some 380 experimental and 137 control studies reported in 117 publications. The authors concluded that the results of 377 of the experimental studies produced a fixed-effect result 3.6 sigma above chance (p = 0.0002) and a random-effects model 4.1 sigma above chance (p = 2.25 x $10^{-5}$). Three studies were excluded because each contained over a billion random bits, and together those studies amounted to over 210 times as much data as the other 377 studies combined. When just those three studies were analyzed, they resulted in a -4.03 sigma deviation from chance (p = 5.6 x $10^{-5}$, two-tailed). This outcome was opposite to the direction of the assigned intention, but it nevertheless confirmed the presence of an anomaly associated with mental influence of the probabilities of random bits.

Despite these findings, the authors concluded that the results were due to selective reporting biases. A reanalysis subsequently argued that their conclusion was flawed because the selective reporting model that the authors used could not account for the observed heterogeneity of the experimental results.[85] That argument was later confirmed by independent analysts.[86]

*From mental intention to attention*

A second class of RNG experiments tested the hypothesis that it was not necessary to apply *intention* to shift the probabilities of random bits, but just focused *attention* in the vicinity of RNGs would also produce negentropic deviations.[87] Dubbed experiments in "field consciousness," these studies were devised as a way to test the assumption that if mind and matter were intimately correlated in some way, then during periods when the attention of many minds became coherent, then aspects of the physical environment might reflect that coherence by also displaying coherent behavior. Many small-scale studies were conducted during meditation gatherings, choral groups, theatrical presentations, and so on;[88–91] most showed the predicted effects.

The single most ambitious field consciousness experiment began in 1998 by placing RNGs in major cities around the world. Each RNG generated a random sample of the sum of 200 bits per second, then uploaded those samples over the Internet every 5 minutes to a server in Princeton, NJ. The study was designed to run continuously as a passive monitor of fluctuations in the RNG network during major world events. The events of interest were inferred to cause periods of coherent mental attention in millions of people, driven by live worldwide media coverage. The events included political elections, terrorist attacks, damaging earthquakes, funerals of beloved celebrities, the opening ceremonies of the Olympics, and so on.

The planned analysis for each event, which always took place before examining the data, involved determining how long the event lasted (generally 1 to 6 hours), and then retrieving the data and calculating the combined sample variance produced by all of the RNGs running during that time period.[92] The formal experiment consisted of 500 such events, which was achieved in 2015. At that point, the overall outcome was a 7.3 sigma effect ($p = 1.43 \times 10^{-13}$). Because the results and implications of this experiment were so remarkable, the raw data (available at global-mind.org) and the reported analyses were double-checked by a third party and were confirmed to be correct.[93]

For nearly a century studies exploring psychokinetic effects on random physical systems have provided evidence that the probability of truly random events can be influenced by the mind. The effects in individual experiments are generally small and variable, but the accumulated data indicate a genuine effect. When these studies first began, it was assumed that mental intention would independently influence each random event. If that were the case, then the experimental yield could easily be enhanced by just increasing the sample size, say, from an RNG producing 200 bits per second to one producing 2 million bits per second. However, because these effects are not simple physical phenomena, but complex psychophysical interactions involving highly variable psychological components, such straightforward amplifications have not been observed. This puzzling observation has led to other interpretations about how mind and matter might interact. The interpretations range from teleological, to goal-directed, to retrocausal effects.[94–98]

*Mental influence of interference in optical systems*

A more recent class of psychokinetic experiments has focused on von Neumann's proposal that the transition from quantum Possibles into classical Actuals involves a psychophysical component.[22,99] The experiment uses optical double-slit interferometers because it is well accepted (if not well understood) that gaining "which-path" information about the path that photons take as they pass through the slits causes the wave-like interference pattern to shift into a particle-like diffraction pattern.

To perform such a test, individuals are asked to focus their attention toward or away from the double-slit apparatus to try to mentally gain information about the photons' path (which cannot be seen with the naked eye), or alternatively, to intentionally alter their paths. To date, 28 experiments based on this idea have been reported by four labs using different protocols, continuous beam gas and diode lasers, single-photon designs, and analytical approaches.[22–25,100,101] Of these tests, 11 are reportedly statistically significant at $p < 0.05$ (two-tail), where just one would be expected by chance. The exact binomial cumulative probability of the published results so far is 5.2 sigma above chance ($p < 5 \times 10^{-8}$)

To date, commentaries about these studies have appeared in five publications. Sassoli de Bianchia and later Pradhan did not question the reported results, but they did offer different theoretical interpretations.[102,103] Baer reanalyzed data from one of the published studies using a simpler method he devised, [23] and after a statistical correction his analysis confirmed the original results.[104] In another reanalysis, Tremblay examined data from a two-year online double-slit experiment. [25,105] He confirmed that the reported results were correct, but then he decided to analyze each of the two years of data separately. Considering those datasets as two independent experiments resulted in reduced statistical power, and thus the result of each year's analysis taken separately was not significant. In another commentary, Walleczek and von Stillfried claimed that a control condition was a false positive in an as-yet unpublished double-slit study.[106] A subsequent reply to that article showed that the critique was flawed because it failed to adjust the results for multiple comparisons, which was required by the protocol of that experiment.[107]

As a relatively new line of mostly exploratory research, caution is warranted in interpreting these results. But it is worth noting that the outcome observed to date is consistent with the more mature empirical database involving RNGs as psychokinetic targets. If successfully confirmed by future studies, this line of research would strengthen the suggestion that Mind plays a role in the becoming of the universe.

## Discussion

### Responsible Free Will

For the first time since Newton, psi experiments in general and psychokinetic studies in particular could scientifically allow for a "responsible free will." In these experiments, participants "try" to alter the probability of truly random events, or to alter the fringes in an optical interference pattern. "Try" is another word for "Will". In our normal lives, we believe that we convert intended Possibles to Actuals all the time by choosing and doing, where counterfactually we could have chosen and done otherwise.

The core question is this: Can such a free will be responsible? In Quantum Mechanics, the outcome of measurement is ontologically indeterminate, breaking the causal closure of classical physics, but entirely random as given by the squared amplitudes of the wave function via the Born Rule. I can act with apparent freedom, but I will still randomly hit the little old lady on the head with a rock. I behave as though I am free, but if freedom is only apparent, then I am not responsible.

However, if the psychokinetic studies are valid, then a human can not only "try" to alter the outcome of a physical system by intentionally altering the probabilities of the outcomes of measurement, say by bending the Born rule, but their will can actually accomplish their desire. Thus, Mind trying and doing can alter the outcome of "actualization" to behave non-randomly. A responsible free will is not ruled out.

The proposal that mind mediates actualization of potentia contains a deep difficulty: Before, e.g, human minds, measurements nevertheless occurred. How? One answer is a form of panpsychism where interacting quantum variables measure one another. There are grounds to hold this view of Quantum Mechanics. It is consistent with the Strong Free Will Theorem that says that electrons "freely decide" to become Up or Down upon measurement.[108] The now confirmed "dud bomb experiment" demonstrates that quantum variables can act on one another even in "interaction-free" experiments.[109,110]

### The Hard Problem

Our proposal does not answer Chalmer's hard problem of qualia. But we do suggest that a "burst of consciousness" happens upon a quantum actualization. In this we parallel Hameroff and Penrose's proposal [36], but we note that their position links an event of consciousness with collapse of a superposition of multiple potential spacetimes into one actual spacetime. They may be correct, but there is no obvious reason to link their version of quantum gravity with an event of consciousness. We would also note that consciousness plays no role in Hameroff and Penrose's spacetime collapse, so in their model consciousness is not an active agent; it is epiphenomenal. By contrast, our proposal that reality consists in Possibles and Actuals linked by measurement invites a natural place for Mind: it is the means by which quantum potentials are actualized. The empirical results of psychokinetic experiments support this suggestion.

### Conclusion

If the mind-body relationship were accounted for by purely classical physics, then quantum mind is impossible. This belief was taken for granted by many scientists until a few decades ago, which is largely why the existence for psi was also considered to be impossible. Today, with growing evidence for quantum biological effects, the plausibility that some aspects of brain function are quantum is increasing, and as such it is time to take the evidence for quantum mind, and thus psi, more seriously. Given what is coming to light in the first decades of the 21$^{st}$ century, it may not be irresponsible to take evidence of psi as an indication that there are spatial and temporal forms of non-locality in living systems, as well as an active role for mind in the physical world.

Compared to the research on particle physics being conducted at CERN, psi experiments are extremely inexpensive, and yet they carry implications at least as profound as any observed in physics. If further experiments continue to show positive effects, especially if these phenomena are shown to have a genetic component, then we will be more secure in their reality and be driven to ask about the possible survival value such phenomena have played in the long evolution of mind since the origins of life on Earth some 3.7 billion years ago.

Mind, in short, may have had – and still have – a role in the evolution of the world.

### Declarations

The authors state that they have no conflicting interests, and this research did not receive any specific grant from funding agencies in the public, commercial, or not-for-profit sectors.

# Bibliography

1  Fodor J. 1992. The big idea: Can there be a science of mind? Times Lit. Suppl. 5.

2  Chalmers DJ. The conscious mind : in search of a fundamental theory. Oxford University Press, New York. 1996.

3  Tornau C. Saint Augustine. The Stanford Encyclopedia of Philosophy. E N Zalta, ed. Metaphysics Research Lab, Stanford University.

4  Nadler S. Spinoza: A Life. Cambridge University Press, Cambridge ; New York. 2018.

5  Stubenberg L. Neutral Monism. The Stanford Encyclopedia of Philosophy. E N Zalta, ed. Metaphysics Research Lab, Stanford University.

6   Downing L. George Berkeley. The Stanford Encyclopedia of Philosophy. E N Zalta, ed. Metaphysics Research Lab, Stanford University.

7   Dennett D. Consciousness Explained. Little, Brown & Company, New York City. 1991.

8   Tononi G, Boly M, Massimini M, Koch C. 2016. Integrated information theory: from consciousness to its physical substrate. Nat Rev Neurosci. **17**(7): 450.

9   Descartes R. Discourse on Method and Meditations on First Philosophy, 4th Ed. Hackett Publishing Company, Indianapolis. 1999.

10  Newton SI. The Principia: The Authoritative Translation: Mathematical Principles of Natural Philosophy. University of California Press, Oakland. 2016.

11  McKenna M, Coates DJ. Compatibilism. The Stanford Encyclopedia of Philosophy. E N Zalta, ed. Metaphysics Research Lab, Stanford University.

12  Goodman R. William James. The Stanford Encyclopedia of Philosophy. E N Zalta, ed. Metaphysics Research Lab, Stanford University.

13  Russell B. The Analysis of Mind. Macmillan, New York. 1921.

14  Rieper E, Gauger E, Morton JJL, Benjamin SC, Vedral V. 2009. Quantum coherence and entanglement in the avian compass. .

15  Brookes JC. 2017. Quantum effects in biology: golden rule in enzymes, olfaction, photosynthesis and magnetodetection. Proc. R. Soc. Math. Phys. Eng. Sci. **473**(2201): 20160822.

16  Ritz T, Damjanović A, Schulten K. 2002. The quantum physics of photosynthesis. ChemPhysChem. **3**(3): 243.

17  Heisenberg W. Physics and Philosophy: The Revolution in Modern Science. Harper Perennial Modern Classics, New York. 2007.

18  Kauffman SA. Humanity in a Creative Universe. Oxford University Press, New York, NY. 2016.

19  Kastner R, Kauffman S, Epperson M. Taking Heisenberg's Potentia Seriously. Int. J. Quantum Found. **4**(2): 158.

20  Kauffman S. 2020. A New View of Reality: Mind and Actualization of Quantum Potentia. J. Cogn. Sci. **21**(3): 475.

21  Shimony A. On Mentality, Quantum Mechanics, and the Actualization of Potentialities The Large, the Small and the Human Mind,. Cambridge University Press. pp 144–60.

22  Radin DI, Michel L, Galdamez K, Wendland P, Rickenbach R, Delorme A. 2012. Consciousness and the double-slit interference pattern: Six experiments. Phys. Essays. **25**(2): .

23  Radin D, Johnston, J., Delorme A. 2013. Psychophysical interactions with a double-slit interference pattern. Phys. Essays. **26**(4): 553.

24  Radin D, Michel L, Pierce A, Delorme A. 2015. Psychophysical interactions with a single-photon double-slit optical system. Quantum Biosyst. **6**(1): 82.

25  Radin D, Michel L, Delorme A. 2016. Psychophysical modulation of fringe visibility in a distant double-slit optical system. Phys. Essays. **29**(1): 14.

26  Aspect A, Dalibard J, Roger G. 1982. Experimental Test of Bell's Inequalities Using Time-Varying Analyzers. Phys. Rev. Lett. **49**(25): 1804.

27  Filk T. 2013. Temporal Non-locality. Found. Phys. **43**(4): 533.

28  Megidish E, Halevy A, Shacham T, Dvir T, Dovrat L, Eisenberg HS. 2013. Entanglement Swapping between Photons that have Never Coexisted. Phys. Rev. Lett. **110**(21): 210403.

29  Radin D. Real Magic: Ancient Wisdom, Modern Science, and a Guide to the Secret Power of the Universe. Harmony, New York. 2018.

30  Wahbeh H, Radin D, Mossbridge J, Vieten C, Delorme A. 2018. Exceptional experiences reported by scientists and engineers. EXPLORE. **14**(5): 329.

31  Carroll S. 2008. Telekinesis and Quantum Field Theory. Sean Carroll. https://www.preposterousuniverse.com/blog/2008/02/18/telekinesis-and-quantum-field-theory/.

32  Cardeña E. 2018. The experimental evidence for parapsychological phenomena: A review. Am. Psychol. **73**(5): 663.

33  Utts J. 2016. Appreciating Statistics. J. Am. Stat. Assoc. **111**(516): 1373.

34  Reber AS, Alcock JE. 2019. Searching for the impossible: Parapsychology's elusive quest. Am. Psychol. No Pagination Specified.

35  Sbitnev VI. 2016. Quantum consciousness in warm, wet and noisy brain. Mod. Phys. Lett. B. **30**(28): 1650329.

36  Hameroff S, Penrose R. 2014. Consciousness in the universe: A review of the 'Orch OR' theory. Phys. Life Rev. **11**(1): 39.

37  Jahn RG, Dunne BJ. 1986. On the quantum mechanics of consciousness with application to anomalous phenomena. Found. Phys. **16**: 721.

38  Josephson BD, Pallikari-Viras F. 1991. Biological utilization of quantum nonlocality. Found. Phys. **21**(2): 197.

39  Mattuck R. 1982. A crude model of the mind-matter interaction using Bohm-Bub hidden variables. J. Soc. Psych. Res. **51**(790): 238.

40  Chari CTK. 1972. Precognition, probability, and quantum mechanics. J. Am. Soc. Psych. Res. J. Am. Soc. Psych. Res. **66**(2): 193.

41  Walker EH. 1979. The quantum theory of psi phenomena. Psychoenergetic Syst. **3**: 259.

42  Squires EJ. Conscious mind in the physical world. A. Hilger, Bristol, England ; New York, NY. 1990.

43  Bem DJ, Honorton C. 1994. Does Psi Exist? Replicable Evidence for an Anomalous Process of Information Transfer. Psychol. Bull. **115**: 4.

44  Dalton KS, Morris RL, Delanoy DL, Radin DI, Taylor R, Wiseman R. 1996. Security Measures in an Automated Ganzfeld System. J. Parapsychol. **60**(2): 129.

45  Utts J. 1991. Replication and meta-analysis in parapsychology. Stat. Sci. 363.

46  Storm L, Tressoldi PE, Di Risio L. 2010. Meta-analysis of free-response studies, 1992-2008: assessing the noise reduction model in parapsychology. Psychol. Bull. **136**(4): 471.

47  Storm L, Tressoldi P. in press. Meta-analysis of free-response studies 2009-2018: Assessing the noise-reduction model ten years on. J. Soc. Psych. Res. .

48  Jarosz A, Wiley J. 2014. What Are the Odds? A Practical Guide to Computing and Reporting Bayes Factors. J. Probl. Solving. **7**(1): .

49  H. Jeffreys. The Theory of Probability. Oxford, Oxford. 1961.

50  Lamb E. 2012. 5 Sigma What's That? Sci. Am. Blog Netw. https://blogs.scientificamerican.com/observations/five-sigmawhats-that/.

51  Vieten C et al. 2018. Future directions in meditation research: Recommendations for expanding the field of contemplative science. PLOS ONE. **13**(11): e0205740.

52  Thalbourne MA. 2000. Transliminality and creativity. J. Creat. Behav. **34**(3): 193.

53  Storm L, Tressoldi P. 2017. Gathering in more sheep and goats: a meta-analysis of forced-choice sheep-goat ESP studies, 1994-2015. J. Soc. Psych. Res. **81**(2): 79.

54  Baptista J, Derakhshani M. 2014. Beyond the coin toss: Examining Wiseman's criticisms of parapsychology. J. Parapsychol. **78**(1): 56.

55  Cohn SA. 1999. Second sight and family history: pedigree and segregation analyses. J. Sci. Explor. **13**(3): 351.

56  Wahbeh H, Radin D, Yount G, Woodley of Menie M, Sarraf M, Karpuj M. in press. Genetics of Psychic Ability - A Pilot Case-Control Exome Sequencing Study. Explore. .

57  Delgado-Romero EA, Howard GS. 2005. Finding and Correcting Flawed Research Literatures. Humanist. Psychol. **33**(4): 293.

58  Duane TD, Behrendt T. 1965. Extrasensory electroencephalographic induction between identical twins. Science. **150**(3694): 367.

59  Targ R, Puthoff H. 1974. Information transmission under conditions of sensory shielding. Nature. **251**: 602.

60  Standish LJ, Johnson LC, Kozak L, Richards T. 2003. Evidence of correlated functional magnetic resonance imaging signals between distant human brains. Altern. Ther. **9**: 122.

61  Richards TL, Kozak L, Johnson LC, Standish LJ. 2005. Replicable functional magnetic resonance imaging evidence of correlated brain signals between physically and sensory isolated subjects. J. Altern. Complement. Med. **11**(6): 955.

62  Karavasilis E, Christidi F, Platoni K, Ferentinos P, Kelekis NL, Efstathopoulos EP. 2018. Functional MRI Study to Examine Possible Emotional Connectedness in Identical Twins: A Case Study. EXPLORE. **14**(1): 86.

63  Radin D, Pierce A. Psi and Psychophysiology Parapsychology: A Handbook for the 21st Century. McFarland, North Carolina, USA.

64  Jensen CG, Parker A. 2012. Entangled in the Womb? A Pilot Study on the Possible Physiological Connectedness Between Identical Twins with Different Embryonic Backgrounds. EXPLORE. **8**(6): 339.

65  Parker A, Jensen C. 2013. Further Possible Physiological Connectedness Between Identical Twins: The London Study. EXPLORE. **9**(1): 26.

66  Schmidt S, Schneider R, Utts J, Walach H. 2004. Distant intentionality and the feeling of being stared at: two meta-analyses. Br J Psychol. **95**(Pt 2): 235.

67  Honorton C, Ferrari DC. 1989. "Future telling": A meta-analysis of forced-choice precognition experiments, 1935-1987. J. Parapsychol. **53**: 281.

68  Radin DI. 1997. Unconscious perception of future emotions: An experiment in presentiment. J. Sci. Explor. **11**: 163.

69  Mossbridge J, Radin D. 2018. Precognition as a form of prospection: A review of the evidence. Psychol. Conscious. Theory Res. Pract. **5**(1): 78.

70  Mossbridge J, Tressoldi P, Utts J. 2012. Predictive physiological anticipation preceding seemingly unpredictable stimuli: a meta-analysis. Front. Psychol. **3**: .

71  Duggan M, Tressoldi P. 2018. Predictive physiological anticipatory activity preceding seemingly unpredictable stimuli: An update of Mossbridge et al's meta-analysis. F1000Research. **7**: 407.

72  Yagi Y, Inoue K. 2018. The Contribution of Attention to the Mere Exposure Effect for Parts of Advertising Images. Front. Psychol. **9**: .

73  Pugnaghi G, Memmert D, Kreitz C. 2019. Examining effects of preconscious mere exposure: An inattentional blindness approach. Conscious. Cogn. **75**: 102825.

74  Bem DJ. 2011. Feeling the future: Experimental evidence for anomalous retroactive influences on cognition and affect. J. Pers. Soc. Psychol. **100**(3): 407.

75  Bem D, Tressoldi P, Rabeyron T, Duggan M. 2015. Feeling the future: A meta-analysis of 90 experiments on the anomalous anticipation of random future events. F1000Research. **4**: 1188.

76  Bacon F. Sylva Sylvarum or a Naturall Historie. William Rawley. 1639.

77  Rhine JB. 1943. Dice thrown by cup and machine in PK tests. J. Parapsychol. **7**(3): 207.

78  Radin DI, Ferrari DC. 1991. Effects of consciousness on the fall of dice: A meta-analysis. J. Sci. Explor. **5**(1): 61.

79  Schmidt H. 1970. A quantum mechanical random number generator for psi tests . J. Parapsychol. **34**(3): 219.

80  Schmidt H. 1987. The strange properties of psychokinesis. J. Sci. Explor. **1**(2): 103.

81  JM Houtkooper. 2002. Arguing for an Observational Theory of Paranormal Phenomena. J. Sci. Explor. **16**(2): 171.

82  Jahn RG, Dunne BJ, Nelson RG, Dobyns YH, Bradish GJ. 2007. Correlations of random binary sequences with pre-stated operator intention: a review of a 12-year program. Explore N. Y. N. **3**(3): 244.

83  Radin D, Nelson R. 1989. Evidence for consciousness-related anomalies in random physical systems. Found. Phys. **19**(12): 1499.

84  Bosch H, Steinkamp F, Boller E. 2006. Examining psychokinesis: the interaction of human intention with random number generators—a meta-analysis. Psychol. Bull. **132**(4): 497.

85  DI Radin, RD Nelson, Y Dobyns, J Houtkooper. 2006. Reexamining psychokinesis: Commentary on the Bösch, Steinkamp and Boller meta-analysis. Psychol. Bull. (529–532): .

86  Varvoglis M, Bancel PA. Micro-psychokinesis Parapsychology: A handbook for the 21st century. McFarland & Co, Jefferson, NC, US. pp 266–81.

87  R. D. NELSON RGJ. 1998. FieldREG II: Consciousness Field Effects: Replications and Explorations. J. Sci. Explor. **12**(3): 425.

88  Radin D, Atwater FH. 2009. Exploratory evidence for correlations between entrained mental coherence and random physical systems. J. Sci. Explor. **23**(3): 263.

89  Radin DI. 2002. Exploring relationships between random physical events and mass human attention: Asking for whom the bell tolls. J. Sci. Explor. **16**(4): 533.

90  Jonas WB, Crawford CC. 2004. The healing presence: Can it be reliably measured? J. Altern. Complement. Med. **10**(5): 751.

91  Dean Radin. The Conscious Universe. HarperOne, New York City. 1997.

92  Bancel P, Nelson R. 2008. The GCP event experiment: Design, analytical methods, results. J. Sci. Explor. **22**(3): 309.

93  Nelson R, Bancel P. 2011. Effects of Mass Consciousness: Changes in Random Data during Global Events. EXPLORE. **7**(6): 373.

94  Bancel PA. 2017. Searching for Global Consciousness: A 17-Year Exploration. EXPLORE. **13**(2): 94.

95  May E, Utts J, Spottiswoode SJP. 1995. Decision augmentation theory: Toward a model of anomalous mental phenomena. J. Parapsychol. **59**: 195.

96  Radin D. 2006. Experiments testing models of mind-matter interaction. J. Sci. Explor. **20**(3): 375.

97  Schmidt H. 1984. Comparison of a teleological model with a quantum collapse model of psi. J. Parapsychol. **48**(4): 261.

98  Varvoglis MP. 1986. Goal-directed- and observer-dependent PK: An evaluation of the conformance-behavior model and the observational theories . J. Am. Soc. Psych. Res. **80**(2): 137.

99  von Neumann J. Mathematical Foundations of Quantum Mechanics. Princeton University Press, Princeton, NJ. 1955.

100     Ibison M, Jeffers S. 1998. A double-slit diffraction experiment to investigate claims of consciousness-related anomalies. J. Sci. Explor. **12**: 543.

101     Guerrer G. 2019. Consciousness-related interactions in a double-slit optical system. https://osf.io/qdkvx/. .

102     Sassoli de Bianchi M. 2013. Quantum measurements are physical processes. Comment on "Consciousness and the double-slit interference pattern: Six experiments," by Dean Radin et al. [Phys. Essays 25, 157 (2012)]. Phys. Essays. **26**(1): 15.

103     Pradhan RK. 2015. An explanation of psychophysical interactions in the quantum double-slit experiment - Physics Essays Publication. Phys. Essays. **28**(3): 324.

104     Radin D, Michel L, Delorme A. 2015. Reassessment of an independent verification of psychophysical interactions with a double-slit interference pattern. Phys. Essays. **28**(4): 415.

105     Tremblay N. 2019. Independent re-analysis of alleged mind-matter interaction in double-slit experimental data. PLOS ONE. **14**(2): e0211511.

106     Walleczek J, von Stillfried N. 2019. False-Positive Effect in the Radin Double-Slit Experiment on Observer Consciousness as Determined With the Advanced Meta-Experimental Protocol. Front. Psychol. **10**: .

107     Radin D, Wahbeh H, Michel L, Delorme A. 2020. Commentary: False-Positive Effect in the Radin Double-Slit Experiment on Observer Consciousness as Determined With the Advanced Meta-Experimental Protocol. Front. Psychol. **11**: .

108     Conway J, Kochen S. 2009. The strong free will theorem. Not. Amer Math Soc. **56**(2): 226.

109     Elitzur AC, Vaidman L. 1993. Quantum mechanical interaction-free measurements. Found. Phys. **23**(7): 987.

110     Kwiat PG, Weinfurter H, Zeilinger A. Interaction-Free Measurement of a Quantum Object: On the Breeding of "Schrödinger Cats" Coherence and Quantum Optics VII. J H Eberly, L Mandel and E Wolf, ed. Springer US, Boston, MA. pp 673–4.